\documentclass[preprint,preprintnumbers,nofootinbib]{revtex4-1}

\usepackage[normalem]{ulem}
\usepackage{graphicx}
\usepackage{dcolumn}
\usepackage{bm}
\usepackage{amsmath}
\usepackage{amsfonts} 
\usepackage{latexsym}
\usepackage{bbm}
\usepackage{color}
\usepackage{amssymb}
\usepackage{amsthm}
\usepackage{epsfig}
\usepackage{physics}
\usepackage{hyperref}
\usepackage{comment}
\usepackage{ulem}
\usepackage{tikz-feynhand}
\hypersetup{
setpagesize=false,
 bookmarksnumbered=true,
 bookmarksopen=true,
 colorlinks=true,
 linkcolor=blue,
 citecolor=red,
}

\begin{document}

\preprint{\textbf{OCHA-PP-378}}

\title{A complex singlet extension of the Standard Model with a singlet fermion dark matter}

\author{Gi-Chol Cho$^1$}
\email{cho.gichol@ocha.ac.jp}

\author{Chikako Idegawa$^1$}
\email{c.idegawa@hep.phys.ocha.ac.jp}

\author{Rie Inumiya$^2$}
\email{rie.inumiya@hep.phys.ocha.ac.jp}

\affiliation{$^1$Department of Physics, Ochanomizu University, Tokyo 112-8610, Japan}

\affiliation{$^2$Graduate School of Humanities and Sciences, Ochanomizu University, Tokyo 112-8610, Japan}

\bigskip

\date{\today}
%chog

\begin{abstract}
We examine a complex singlet scalar extension of the Standard Model (CxSM) with an extra singlet fermion. Both the singlet scalar and fermion are dark matter (DM) candidates.
It is known that although the scalar potential in the CxSM can realize strong first-order electroweak phase transition, the scalar DM included in the model gives only a tiny amount of the relic density compared to the observed one. Therefore, a fermion DM is introduced to compensate for the lack of relic density. 
We find that the scattering of the fermion DM and nucleons is sufficiently suppressed when the masses of scalar mediators are degenerate, as is the case of the scalar DM. We show the range of a combination of the mass and the Yukawa coupling of the fermion DM, which satisfies both the observed relic density and conditions of strong first-order electroweak phase transition. 
\end{abstract}

\maketitle

\section{Introduction}\label{sec:intro}

The existence of dark matter (DM), suggested by numerous cosmological observations, provides direct evidence for physics beyond the Standard Model (SM). Among the many possible DM candidates whose properties are consistent with observations, the weekly interacting massive particle (WIMP) is an attractive candidate that can explain DM abundance thermally. However, contrary to the efforts of various DM search experiments, including high-energy accelerator experiments and DM direct detection experiments, no DM signal has yet been discovered. In particular, the recent LUX-ZEPLIN (LZ) experiment~\cite{LZ:2022ufs} has placed strong upper bounds on the spin-independent scattering cross section between WIMP-DM and nucleons. Therefore, it is important to construct a DM model that is consistent with the severe constraint.

A class of DM models that have been proposed to suppress the scattering of DM and nucleons is the so-called pseudo Nambu-Goldstone (pNG) DM model~\cite{Gross:2017dan, Jiang:2019soj, Abe:2021nih, Liu:2022evb} stems from the complex singlet scalar extension of the Standard Model (CxSM)~\cite{Barger:2008jx}. This model extends the SM by adding a complex singlet scalar field $S$. The imaginary part of $S$ behaves as a DM whose stability is ensured by the CP symmetry of the scalar potential. On the other hand, the real part of $S$ mixes with the SM Higgs boson to form the mass eigenstates $h_1$ and $h_2$, which mediate scattering between DM and quarks.

 The CxSM imposing the global U(1) symmetry on the scalar potential, which is softly broken by an operator with mass dimension two, has been well studied~\cite{Gross:2017dan, Jiang:2019soj,  Liu:2022evb}. The scattering amplitudes of DM and nucleons with $h_1$ and $h_2$ as mediator particles in this model are suppressed in the limit of zero momentum transfer~\cite{Gross:2017dan}. However, this model suffers from the domain-wall problem due to spontaneous $Z_2$ symmetry breaking when $S$ develops the vacuum expectation value (VEV). The authors of ref.~\cite{Abe:2021nih} proposed to extend the scalar potential by introducing a linear term of $S$ to the scalar potential to solve this problem\footnote{Another possibility to avoid the domain-wall problem has been proposed in ref.~\cite{Abe_2023}, by requiring a complex scalar field transforms under a global SU(2) and an extra U(1) gauge symmetries.}. In this model, the scattering cross section is suppressed when the masses of $h_1$ and $h_2$ are nearly degenerate. This suppression mechanism is called a degenerate scalar scenario.

Generally, large DM mass and/or small interactions with SM particles are preferred to explain direct detection results. On the other hand, unlike decoupling the DM from the SM, the degenerate scalar scenario employs the built-in mechanism to suppress DM signals, independent of DM masses and interactions. Therefore, this study focuses on the Higgs degenerate region, which is expected to highlight blind spots in previous experiments and provide directions for future experiments.

The origin of baryon asymmetry of the Universe (BAU) is also one of the open questions in particle physics and cosmology. The most testable scenario to realize BAU is electroweak baryogenesis (EWBG)~\cite{Kuzmin:1985mm, Rubakov:1996vz, Funakubo:1996dw, Riotto:1998bt, Trodden:1998ym, Bernreuther:2002uj, Cline:2006ts, Morrissey:2012db, Konstandin:2013caa, Senaha:2020mop} associated with strong first-order electroweak phase transition (EWPT)~\cite{Kajantie:1996mn, Rummukainen:1998as, Csikor:1998eu, Aoki:1999fi}, which implies new physics beyond the SM. The authors of ref.~\cite{Cho:2021itv} have stated that introducing $S$ is also effective in realizing EWPT. In addition, they have pointed out conditions on the model parameters for the degenerate scalar scenario and one for strong first-order EWPT conflict. 
These two conditions are satisfied simultaneously only when the DM mass is half of the Higgs boson mass, at which point the DM relic density is much lower than the observed value $\Omega_\mathrm{DM}h^2 = 0.1200 \pm 0.0012$~\cite{Planck:2018vyg}. This requires another DM candidate to explain the observed relic density.

In this paper, we introduce a singlet fermion to the CxSM as another DM candidate, which compensates for the lack of relic density via the scalar DM in the CxSM\footnote{An extension of the CxSM introducing a singlet fermion has been studied extensively in ref.~\cite{Chen:2022vac}. However, the work focuses on the case where the scalar potential does not have CP symmetry, so the pNG boson is no longer stable. As a result, the singlet fermion plays as the only DM candidate.}. We show that, without conflicting with the current constraints from the direct detection experiments, there is allowed model parameter space, which explains the observed relic density of the DM and is favored by strong first-order EWPT simultaneously. 
In the case of the scalar DM, the suppression mechanism in the DM-quark scatterings requires a particular relation among couplings in the scalar potential~\cite{Cho:2023hek}. The suppression of the scattering processes between the fermion DM and quarks is, however, due to the orthogonality of the mixing matrix, which transforms the current eigenstates of the SM Higgs boson and the CP-even scalar in $S$ to the mass eigenstates $h_1$ and $h_2$. Thus, suppressing the fermion DM and quarks is valid independently from the relation among the couplings in the scalar potential.

The pNG DM models with a singlet fermion DM have been proposed so far~\cite{
DiazSaez:2021pmg, DiazSaez:2023wli, Yaguna:2021rds}.  
The distinctive features of our study compared with those works are as follows: First, we adopt the degenerate scalar scenario to suppress the DM-quark scattering processes so that the model is less constrained from the direct detection experiments. 
Second, we take into account the condition of the model parameter space in realizing the strong first-order EWPT, which has yet to be considered in previous works.

The paper is organized as follows. In Sec.~\ref{sec:model}, we briefly review our model. In Sec.~\ref{sec:dege}, we present cancellation conditions for the scattering between the fermion DM and quarks. Constraints on the parameter space of the fermion DM from the relic density and the direct detection experiments are discussed in Sec.~\ref{sec:pheno}. Sec.~\ref{sec:sum} is devoted to summarizing our study.

\section{The Model}\label{sec:model}

In this section, we briefly review the model, an extension of the CxSM with a singlet fermion, to fix our notation.  
The CxSM extends the SM by adding the complex SU (2) gauge singlet scalar field $S$~\cite{Barger:2008jx}. The following scalar potentials are employed in this study:
\begin{equation}
V(H,S)= \frac{m^2}{2} H^{\dagger} H + \frac{\lambda}{4} (H^{\dagger} H )^2 +\frac{\delta_2}{2} H^{\dagger} H \left| S\right|^2 + \frac{b_2}{2} \left| S\right|^2 + \frac{d_2}{2} \left| S\right|^4 + \left( a_1 S + \frac{b_1}{4} S^2 + \rm{h.c.} \right) ,
\label{potential}
\end{equation}
where both $a_1$ and $b_1$ break the global U(1) symmetry softly. 
The former ($a_1$) is necessary to avoid unwanted $Z_2$ symmetry ($S\to -S$) that could cause domain-wall problem~\cite{Zeldovich:1974uw}, while the latter ($b_1$) needs to make the Nambu-Goldstone boson massive after spontaneous U(1) symmetry breaking.  
Although there could be other soft U(1) breaking terms in the scalar potential, we adopt \eqref{potential} as a minimal and renormalizable model. Furthermore, as will be discussed in Sec.~\ref{sec:dege}, a suppression mechanism of the scattering between the scalar DM and quark prohibits the SU (2)$_L$ doublet-singlet mixing such as $H^{\dagger} H S$ and $H^{\dagger} H S^2$.

The two scalar fields are parametrized as
\begin{align}
H = \frac{1}{\sqrt{2}} \begin{pmatrix} 0 \\ v + h \end{pmatrix}, ~~S =  \frac{1}{\sqrt{2}} \left( v_s + s + i \chi \right),
\label{sca}
\end{align}
where the unitary gauge is employed so that the Goldstone field in $H$ is suppressed. $v~(\simeq 246~\rm{GeV})$ and $v_s$ are VEVs of $H$ and $S$, respectively. In this study, since all the couplings in \eqref{potential} are assumed to be real, the scalar potential is invariant under CP-transformation ($S\to S^*$). Thus, the real and imaginary parts of $S$ do not mix, and the stability of $\chi$ is guaranteed, making it a DM candidate.

First derivatives of the scalar potential concerning $h$ and $s$ are respectively given by 
\begin{align}
\frac{1}{v} \left< \frac{\partial V}{\partial h} \right>
&=
\frac{m^2}{2} + \frac{\lambda}{4} v^2 + \frac{\delta_2}{4} v^2_s = 0 ,
\label{tadpole1}
\\
\frac{1}{v_s} \left< \frac{\partial V}{\partial s} \right>
&=
\frac{b_2}{2} + \frac{\delta_2}{4} v^2 + \frac{d_2}{4} v_s^2 + \frac{\sqrt{2} a_1}{v_s} + \frac{b_1}{2} = 0,
\label{tadpole2}
\end{align}
where $\langle\cdots\rangle$ denotes that all fluctuation fields are zero. Note that $a_1\neq 0$ results in nonzero $v_S$.

The mass matrix of the CP-even states ($h,s$) is represented as
\begin{align}
\mathcal{M}_S^2 
= 
\begin{pmatrix} 
\frac{\lambda}{2} v^2  &  \frac{\delta_2}{2} v v_s \\ \frac{\delta_2}{2} v v_s & \Lambda^2 
\end{pmatrix},
\quad\Lambda^2 \equiv \frac{d_2}{2} v_S^2-\sqrt{2}\frac{a_1}{v_S},
\label{sca2}
\end{align}
which is diagonalized by an orthogonal matrix $O(\alpha)$ as
\begin{align}
O(\alpha)^\top \mathcal{M}_s^2 O(\alpha) =  \begin{pmatrix} m_{h_1}^2 & 0 \\ 0 & m_{h_2}^2 \end{pmatrix},  ~~ O(\alpha) =  \begin{pmatrix} \cos{\alpha} & - \sin{\alpha} \\  \sin{\alpha} & \cos{\alpha} 
\end{pmatrix} ,
\label{mass}
\end{align}
where $\alpha$ is a mixing angle such that $(h, s)^\top = O(\alpha) (h_1,h_2)^\top$ and $\alpha\to0$
corresponds to the SM-like limit. The mass eigenvalues are given by
\begin{align}
m^2_{h_1,h_2} = \frac{1}{2} \left( \frac{\lambda}{2} v^2 + \Lambda^2 \mp \sqrt{\left( \frac{\lambda}{2} v^2 - \Lambda^2 \right)^2 + 4 \left( \frac{\delta_2}{2} v v_s \right)^2 } \right) , 
\label{eigenvalue}
\end{align}
where $h_1$ is assumed that the SM Higgs boson observed at the LHC experiments with $m_{h_1} = 125~\rm{GeV}$~\cite{CMS:2012qbp}. Using the soft breaking terms $a_1$ and $b_1$, the mass of CP-odd state $\chi$ is written as
\begin{align}
m^2_{\chi} = \frac{b_2}{2} - \frac{b_1}{2} + \frac{\delta_2}{4} v^2 + \frac{d_2}{4} v_s^2 = - \sqrt{2} \frac{a_1}{v_s} -b_1 ,
\label{DMmass}
\end{align}
where the tadpole condition (\ref{tadpole2}) is used in the second equality.

Next, we introduce a singlet vector-like fermion field $f$ that couples only to the singlet scalar $S$. The Yukawa interaction between $f$ and $S$ can be expressed as follows 
\begin{align}
- {\mathcal{L}}_f 
&= 
\lambda_f \bar{f} f  S + \rm{h.c.},
\label{fDMyukawa1}
\end{align}
while that of the fermion $f$ and the CP-even scalar $h_1$ or $h_2$ is clearly given by
\begin{align}
- {\mathcal{L}}_f 
&=\frac{\lambda_f}{\sqrt{2}} \bar{f} f \left(h_1 \sin \alpha+h_2 \cos \alpha\right),
\label{fDMyukawa2} 
\end{align}
where $\lambda_f$ is the Yukawa coupling constant and note that $\lambda_f$ and the fermion mass $m_f$ are treated here as independent parameters.

On the other hand, the interaction Lagrangian of a quark $q$ to the CP-even scalar $h_1$ or $h_2$ is written as
\begin{align}
- {\mathcal{L}}_{Y} 
&=\frac{m_q}{v}  \bar{q} q \left( h_1 \cos{\alpha} - h_2 \sin{\alpha} \right) ,
\label{SMqyukawa}
\end{align}
where $m_q$ denotes the mass of the quark $q$.

Theoretical constraints on the model parameters are summarized as follows. A requirement on the scalar potential bounded from below is given by\footnote{If $\delta_2<0$, $\lambda d_2>\delta_2^2$ is also required. In this paper, $\delta_2$ is assumed to be positive.}
\begin{align}
\lambda>0,\quad d_2 > 0.
\label{bfb}
\end{align}
$\lambda$ and $d_2$ are also constraint from perturbative unitarity~\cite{Chen:2014ask} as in
\begin{align}
\lambda < \frac{16 \pi}{3},~~ d_2 < \frac{16 \pi}{3}.
\label{ptbtv}
\end{align}
In addition, by requiring that the eigenvalues of the mass matrix be positive, the stability conditions for the tree-level potential are~\cite{Barger:2008jx}
\begin{equation}
\lambda \left( d_2 - \frac{2 \sqrt{2} a_1}{v_s^3} \right) > \delta_2^2.
\label{stability}
\end{equation}

For later convenience, the relationship between parameters should be mentioned. The original Lagrangian parameters in the scalar potential $\{ m^2, b_2, \lambda, d_2, \delta_2, b_1 \} $ are replaced by $\{ v, v_s, m_{h_1}, m_{h_2}, \alpha, m_{\chi} \}$ while  retaining $a_1$ as an input. $m^2$ and $b_2$ are eliminated from the tadpole conditions (\ref{tadpole1}) and (\ref{tadpole2}) such that
\begin{align}
m^2
&=
-\frac{\lambda}{2}v^2- \frac{\delta_2}{2}v_s^2,
\label{m2}
\\
b_2
&=
- \frac{\delta_2}{2}v^2 - \frac{d_2}{2} v_s^2 - b_1 -2 \sqrt{2} \frac{a_1}{v_s}.
\label{b2}
\end{align}
The remaining four parameters are given by using Eq.~\eqref{mass} and \eqref{DMmass} as follows:
\begin{align}
\lambda
&=
\frac{2}{v^2} \left(m_{h_1}^2 \cos{\alpha}^2 +m_{h_2}^2 \sin{\alpha}^2 \right),
\label{lam}
\\
\delta_2
&=
\frac{1}{v v_s} (m_{h_1}^2 - m_{h_2}^2) \sin{2 \alpha},
\label{del2}
\\
d_2
&=
2 \left( \frac{m_{h_1}^2}{v_s} \right)^2 \sin{\alpha}^2 + 2 \left( \frac{m_{h_2}^2}{v_s} \right)^2 
\cos{\alpha}^2 + 2 \sqrt{2} \frac{a_1}{v_s^3},
\label{d2}
\\
b_1
&=
- m_{\chi}^2 - \sqrt{2} \frac{a_1}{v_s} .
\label{b1}
\end{align}
We also emphasize that the mass of the fermion DM $m_f$ and the Yukawa coupling constant $\lambda_f$ are treated as independent input parameters. Therefore, the overall input parameters are $\{ v, v_s, m_{h_1}, m_{h_2}, \alpha, m_{\chi}, m_f, \lambda_f \}$ and the output parameters are $\{ m^2, b_2, \lambda, d_2, \delta_2, b_1 \} $.

\section{Degenerate scalar scenario}\label{sec:dege}

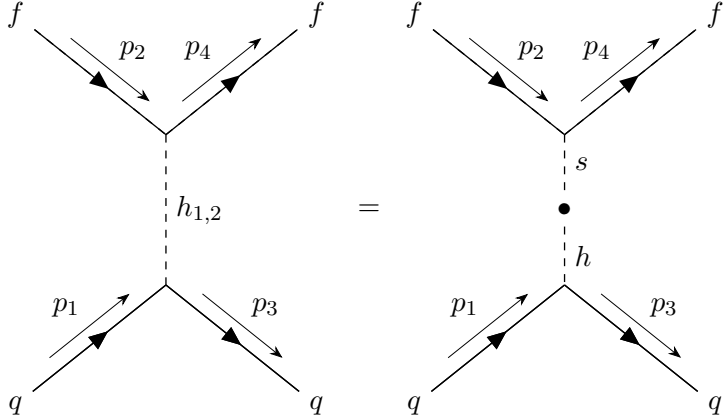
\begin{figure}[htbp]
\centering
    \begin{tabular}{cc}

      \begin{minipage}[t]{0.3\linewidth}
      \begin{tikzpicture}
         \begin{feynhand}
\vertex (a) at (0,0);
\vertex (b) at (-1,0.8);
\vertex (beta) at (-1,1.6);
\vertex (c) at (-1,2.5);
\vertex[particle] (d) at (-2,1.6){$f$};
\vertex[particle] (e) at (2,1.6){$f$};
\vertex (f) at (0,-2);
\vertex[particle] (i) at (2.7,-1) {=};
\vertex[particle] (g) at (-2,-3.6){$q$};
\vertex[particle] (h) at (2,-3.6){$q$};
\propag[fermion] (d) to (a);
\propag[fermion] (a) to (e);
\propag[sca] (a) to [edge label=$h_{1,2}$] (f);
\propag[fermion] (g) to (f);
\propag[fermion] (f) to (h);
\propag[fermion, mom={$p_1$}] (g) to (f);
\propag[fermion, mom={$p_3$}] (f) to (h);
\propag[scalar, mom={$p_2$}] (d) to (a);
\propag[scalar, mom={$p_4$}] (a) to (e);
\end{feynhand}
\end{tikzpicture}
      \end{minipage} &

      \begin{minipage}[t]{0.3\linewidth}
\begin{tikzpicture}
\begin{feynhand}
\vertex (a) at (0,0);
\vertex (b) at (-1,0.8);
\vertex (beta) at (-1,1.6);
\vertex (c) at (-1,2.5);
\vertex[particle] (d) at (-2,1.6){$f$};
\vertex[particle] (e) at (2,1.6){$f$};
\vertex (f) at (0,-2);
\vertex[particle] (i) at (0,-1) {$\bullet$};
\vertex[particle] (g) at (-2,-3.6){$q$};
\vertex[particle] (h) at (2,-3.6){$q$};
\propag[fermion] (d) to (a);
\propag[fermion] (a) to (e);
\propag[sca] (a) to [edge label=$s$] (i);
\propag[sca] (i) to [edge label=$h$] (f);
\propag[fermion] (g) to (f);
\propag[fermion] (f) to (h);
\propag[fermion, mom={$p_1$}] (g) to (f);
\propag[fermion, mom={$p_3$}] (f) to (h);
\propag[scalar, mom={$p_2$}] (d) to (a);
\propag[scalar, mom={$p_4$}] (a) to (e);
\end{feynhand}
\end{tikzpicture}
      \end{minipage}

    \end{tabular}
\caption{Feynman diagrams of the scattering between the vector-like fermion DM $f$ and quark $q$  in the mass eigenstates and the current eigenstates of scalar mediators.}
\label{fig;degenerate}
\end{figure}

 In this section, we discuss a suppression mechanism of the scattering between DM and quark. 
As mentioned previously, both the scalar $\chi$ and the fermion $f$ play a role of DM in our model. 
The scattering of the scalar DM $\chi$ and quark $q$ in this model has already been investigated in ref.~\cite{Abe:2021nih}. In the absence of the linear term of $S$, the scattering amplitude $\mathcal{M}_\chi$ is suppressed by the limit of zero momentum transfer, while nonzero $a_1$ in the scalar potential~\eqref{potential} requires mass degeneracy of two Higgs bosons: 
\begin{align}
 i \mathcal{M}_\chi \simeq i \frac{m_q}{v v_S}\sin\alpha \cos\alpha \bar{u}(p_3) u(p_1) 
\frac{\sqrt{2}a_1}{v_S}\qty(\frac{1}{m_{h_1}^2}-\frac{1}{m_{h_2}^2}),
\label{ampsdm}
\end{align}
where $u(p)~(\bar{u}(p))$ represents an incoming (outgoing) fermion spinor with a momentum $p$. 
However, requirements from the first-order EWPT strongly constrain the parameters to realize the degenerate scalars, which will be discussed in the next section.

Next, we consider the possibility of suppressing the scattering between the fermion DM $f$ and quark, which is 
mediated by $h_1$ and $h_2$, as shown in the diagram on the left in Fig.~\ref{fig;degenerate}. 
It is easy to see that this process can only occur through the mixing between the singlet scalar $s$ and the SM Higgs $h$ 
as shown in the diagram on the right in Fig.~\ref{fig;degenerate} since $f$ only couples to $S$ as can be seen from Eq.~\eqref{fDMyukawa1}.
The scattering amplitude of the fermion DM $\mathcal{M}_f$ is the sum of $\mathcal{M}_1$ and $\mathcal{M}_2$ mediated by $h_1$ and $h_2$, respectively. Using Eq.~\eqref{fDMyukawa2} and \eqref{SMqyukawa}, one finds
\begin{align}
 i\mathcal{M}_f
&=
i\qty(\mathcal{M}_1 + \mathcal{M}_2),
\label{msumtr}
\\
 i\mathcal{M}_1
&=
+ i\frac{m_q}{v} \frac{\lambda_f}{t-m^2_{h_1}} \sin{\alpha}\cos{\alpha}~\bar{u}(p_4)~ u(p_2)~\bar{u}(p_3)~ u(p_1),
\label{m1tr}
\\
 i\mathcal{M}_2
&=
- i\frac{m_q}{v} \frac{\lambda_f}{t-m^2_{h_2}} \sin{\alpha}\cos{\alpha}~\bar{u}(p_4)~ u(p_2)~\bar{u}(p_3)~ u(p_1),
\label{m2tr}
\end{align}
where $t$ describes a momentum transfer, $t\equiv \qty(p_1 - p_3)^2$.

Since the momentum transfer of the direct detection experiment is very small compared to masses of the mediator particle, i.e., $t\ll m_{h_1}^2,m_{h_2}^2$, the scattering amplitude can be expressed as
\begin{align}
 i\mathcal{M}
 =-i\frac{m_q \lambda_f}{v}  \sin{\alpha} \cos{\alpha} ~\bar{u}(p_4)~ u(p_2)~\bar{u}(p_3)~ u(p_1) \left( \frac{1}{m^2_{h_1}}-\frac{1}{m^2_{h_2}} \right) .
\label{msumtr2}
\end{align}
Eq.~\eqref{msumtr2} shows that $m_{h_1}=m_{h_2}$ is required for the two scattering amplitudes to cancel each other. Note that if the two Higgs masses are degenerate, the scattering would be suppressed regardless of adopting the zero momentum transfer limit. Furthermore, we should mention the origin of the degenerate scalar scenario. The degenerate scalar scenario comes from the orthogonality of the mixing matrix
$O(\alpha)$~\eqref{mass}. In addition, in the scattering suppression of the scalar DM and quark, the SU(2)$_L$ doublet-singlet mixing in the scalar potential is restricted (see ref.~\cite{Cho:2023hek} for details). 
On the other hand, since the mass and interactions of the fermion DM are independent of the scalar potential, we emphasize that the scattering suppression of the fermion DM and quark is caused purely by the orthogonality of the mixing matrix and is independent of the details of the scalar potentials.

\begin{figure}[htpb]
\begin{center}
\includegraphics[width=9cm]{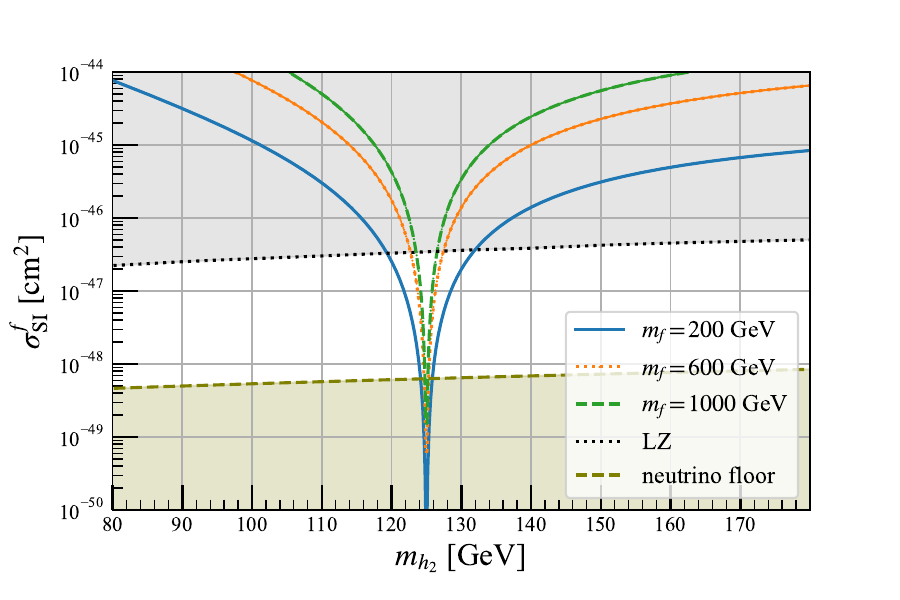}
\caption{Spin-independent cross section of the fermion DM-nucleon scattering $\sigma^f_\mathrm{SI}$ as a function of $m_{h_2}$
for $m_f=$ 200 GeV (blue-solid line), 600 GeV (orange-dotted line), and 1000 GeV (green-dashed line). Other parameters are fixed at $\alpha=\pi/4$ and $v_S=\sqrt[3]{a_1}=246$ GeV. The upper and lower shaded regions are excluded by the LZ experiment~\cite{LZ:2022ufs} and the neutrino floor~\cite{OHare:2021utq}, respectively.}
\label{fig;suppress}
\end{center}
\end{figure}

In Fig.~\ref{fig;suppress}, the spin-independent cross section ($\sigma^f_\mathrm{SI}$) for the scattering between the fermion DM $f$ and nucleons is displayed as a function of $m_{h_2}$. 
The numerical results are obtained using the public code \texttt{micrOMEGAs}~\cite{Belanger:2020gnr}.
To compare with the scalar DM case, we have chosen $\alpha=\pi/4$ and $v_S=\sqrt[3]{a_1}=246$ GeV, faithful to the parameter selection in ref.~\cite{Abe:2021nih}\footnote{Here, values of the parameters are set for just a reference to demonstrate the cancellation mechanism and to compare with previous studies~\cite{Abe:2021nih}. From the next section, numerical considerations will be made with new benchmark points, taking into account the realization of a strong first-order EWPT.} The three lines in Fig.~\ref{fig;suppress} represent $m_f=$ 200 GeV (blue-solid line), 600 GeV (orange-dotted line), and 1000 GeV (green-dashed line), respectively.
The shaded region above the black dotted line and below the yellow dotted line indicate regions that are excluded by the LZ experiment~\cite{LZ:2022ufs} and the background from the elastic neutrino-nucleon scattering (the so-called neutrino floor)~\cite{OHare:2021utq}. 
As expected, there is a large dip around $m_{h_1}\simeq m_{h_2}$ regardless of the size of $m_f$. Thus, the scattering of the fermion DM $f$ and nucleons is suppressed as well as that of the scalar DM $\chi$ and nucleons obtained in ref.~\cite{Abe:2021nih}.

Before proceeding to the next section, we mention that the degenerate scalar scenario is consistent with the bounds from collider experiments on the fermion DM. An invisible decay from the Higgs boson to the fermion DM pair can occur when the fermion DM mass is smaller than half of the Higgs mass. As shown in Eq.~(\ref{fDMyukawa2}), the Yukawa couplings between the fermion DM and $h_1~(h_2)$ are those with the SM Higgs boson multiplied by $\sin{\alpha}~(\cos{\alpha})$. Decay rates from $h_1$ and $h_2$ to the fermion DM is expressed as follows:
\begin{align}
\Gamma_{h_1\to f\bar{f}}&=\sin^2{\alpha}~\Gamma_{h\to f\bar{f}}^{\mathrm{SM}}(m_{h_1}),\\
\Gamma_{h_2\to f\bar{f}}&=\cos^2{\alpha}~\Gamma_{h\to f\bar{f}}^{\mathrm{SM}}(m_{h_2}),
\end{align}
where $\Gamma_{h\to f\bar{f}}^{\mathrm{SM}}(m_{h_{1(2)}})$ is the Higgs partial decay width in the SM as a function of $m_{h_{1(2)}}$. The sum of two processes by $h_1$ and $h_2$ is 
\begin{align}
\Gamma_{h_1\to f\bar{f}}+\Gamma_{h_2\to f\bar{f}} \simeq \Gamma_{h\to f\bar{f}}^{\mathrm{SM}}(m_h), 
\end{align}
holds for any $\alpha$. 
Thus, the invisible decay of the CxSM is the same as that of the SM and is consistent with the bounds from the collider experiment~\cite{ATLAS:2022yvh}\footnote{A similar argument can be made for the Higgs signal strength in the CxSM, which is consistent with that of the SM. See ref.~\cite{Cho:2022zfg} for details.}.

\section{Numerical study}\label{sec:pheno}

In this section, we investigate constraints on the parameter space of the CxSM with the fermion DM from the observed DM relic density and the results of direct detection experiments. 
In the beginning, we briefly review the restriction on the parameters in the scalar potential~\eqref{potential}, realizing the suppression mechanism of the scalar DM-quark scattering confronts strong first-order EWPT.

The scalar DM $\chi$ and nucleons scattering cross section in the limit of vanishing momentum transfer has the following proportional relationship
\begin{align}
\sigma_{\mathrm{SI}}^\chi &\propto \sin ^2 \alpha \cos ^2 \alpha\left(\frac{1}{m_{h_1}^2}-\frac{1}{m_{h_2}^2}\right)^2 \frac{a_1^2}{v_S^4}\nonumber \\
&=
\frac{a_1^2 v^2}{4 m_{h_1}^4 m_{h_2}^4} \frac{\delta_2^2}{v_S^2},
\label{scalarsigma}
\end{align}
where Eq.~\eqref{del2} is used to obtain the second line. 
Therefore, the cancellation mechanism is realized by suppressing $\delta_2$ owing to $m_{h_1}\simeq m_{h_2}$. 
It should be emphasized that, however, even if $\delta_2$ is small, the cancellation mechanism will not work sufficiently for small $v_S$, i.e., $v_S$ should be of a moderate size. 
On the other hand, the requirements on the parameters from strong first-order EWPT conflict with those of the degenerate scalar scenario~\cite{Cho:2021itv}.  
It has been pointed out that the structure of the tree-level scalar potential predominantly induces strong first-order EWPT~\cite{Cho:2021itv}. In the following, we briefly review the constraints imposed on the model parameters by the conditions of strong first-order EWPT, i.e., ${v_C}/{T_C}\gtrsim 1$, where the critical temperature $T_C$ represents the temperature when the potential has two degenerate minima, and the Higgs VEV at that time is written as $v_C$.

The phase transition pattern of our model is 
$\qty(\langle H \rangle, \langle S \rangle)=(0,v_S') \to (v,v_S)$, 
where $\langle H \rangle$ and $\langle S \rangle$ denote the VEVs of the SM Higgs field $H$ and the singlet field $S$, respectively. $T_C$ and $v_C$ can be approximated as follows:\footnote{The $\Sigma_H$ in \eqref{vCTC}
(and $\Sigma_S$ which will appear in \eqref{tdplvsd}) represents the two-point self-energy of $H$ ($S$) and is introduced when qualitatively discussing the effective potential at the finite temperature. 
The explicit expression of $\Sigma_H$ and $\Sigma_S$ can be found in ref.~\cite{Cho:2021itv}. We omit to give them here 
because they do not make a significant contribution to the strong first-order EWPT.}
\begin{align}
v_C \simeq \sqrt{\frac{2 \delta_2\left(v_{S C}'\right)^2}{\lambda}\left(1-\frac{v_{S C}}{v_{S C}'}\right)},\quad T_C \simeq \sqrt{\frac{1}{2 \Sigma_H}\left(-m^2-\frac{\left(v_{S C}'\right)^2}{2} \delta_2\right)},
\label{vCTC}
\end{align}
where quantities evaluated at $T_C$ have a subscript $C$.
To make EWPT first-order, large $\delta_2$ and large $v_{SC}'$ are needed as is clear from Eq.~\eqref{vCTC}.
As can be seen from Eq.~\eqref{del2}, large $\delta_2$ requires small $v_S$ to compensate for the difference in masses of two Higgs bosons. On the other hand, the cubic equation for $v_S'$ is obtained from the tadpole condition of the singlet vacuum $(0,v_S')$:
\begin{align}
\left(v_{S C}'\right)^3+\frac{2\left(b_1+b_2+2 \Sigma_S\right)}{d_2} v_{S C}'+\frac{4 \sqrt{2} a_1}{d_2}=0.
\label{tdplvsd}
\end{align}
When real solutions exist, $v_S'$ is scaled by $1/\sqrt{d_2}$, indicating that small $d_2$ is needed for a large $v_C$. In the degenerate scalar scenario, the representation of $d_2$ \eqref{d2} can be rewritten as
\begin{align}
d_2=\frac{2}{v_S^2}\left[m_{h_1}^2+\left(m_{h_2}^2-m_{h_1}^2\right) \cos ^2 \alpha+\frac{\sqrt{2} a_1}{v_S}\right] \simeq \frac{2}{v_S^2}\left[m_{h_1}^2+\frac{\sqrt{2} a_1}{v_S}\right] .
\end{align}
Positive but small $d_2$ determines the magnitude and the sign of $a_1$, i.e., $a_1<0$.

However, it is easy to see that, from Eq.~\eqref{del2}, the co-existence of sizable $\delta_2$ and small $v_s$ makes suppression of $\sigma_{\mathrm{SI}}^\chi$ weaker significantly. 
It means the cancellation mechanism by degenerate Higgs bosons and the condition realizing strong first-order EWPT conflict~\cite{Cho:2021itv}. 
\begin{figure}[htpb]
\begin{center}
\includegraphics[width=7cm]{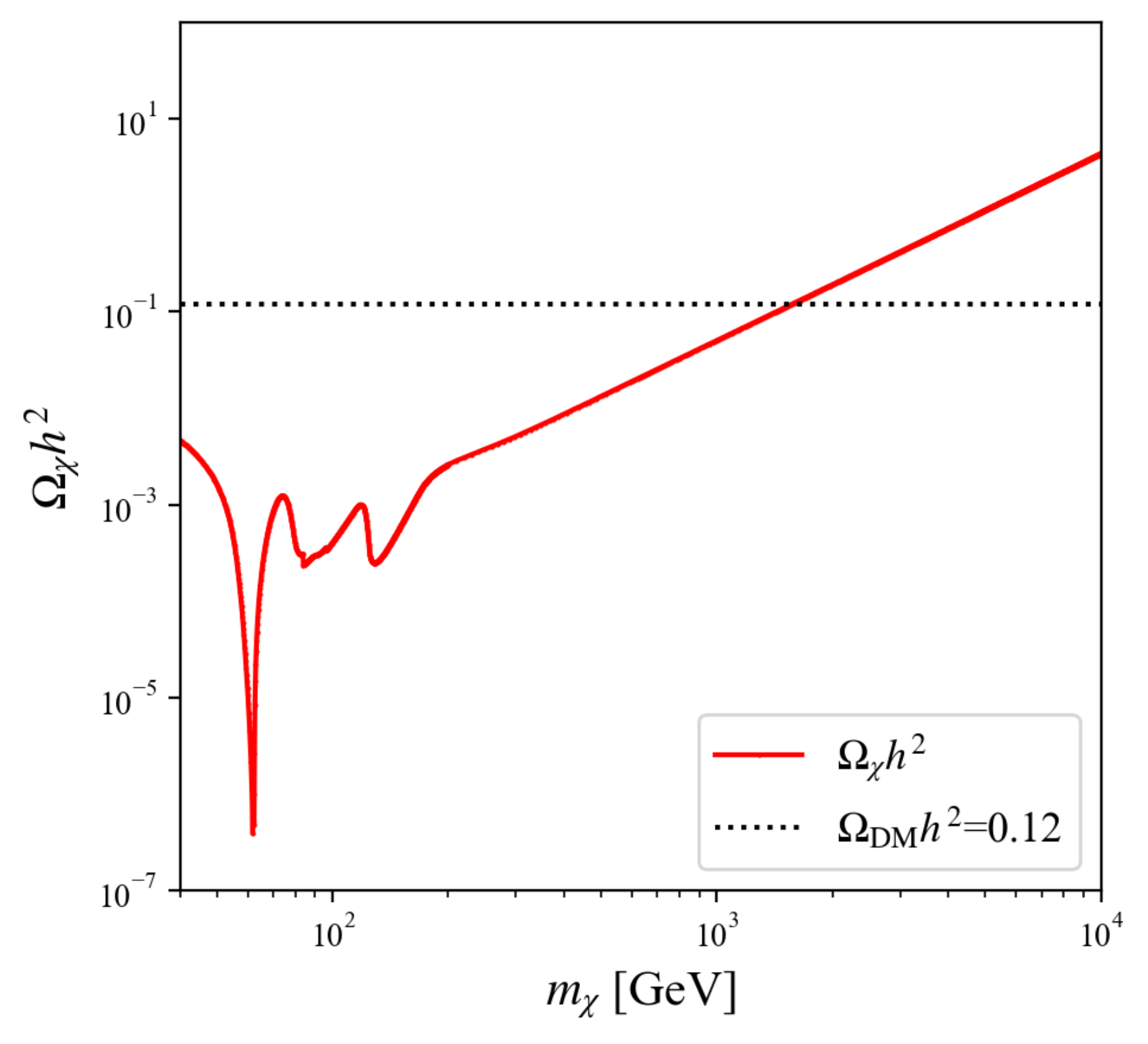}
\includegraphics[width=7cm]{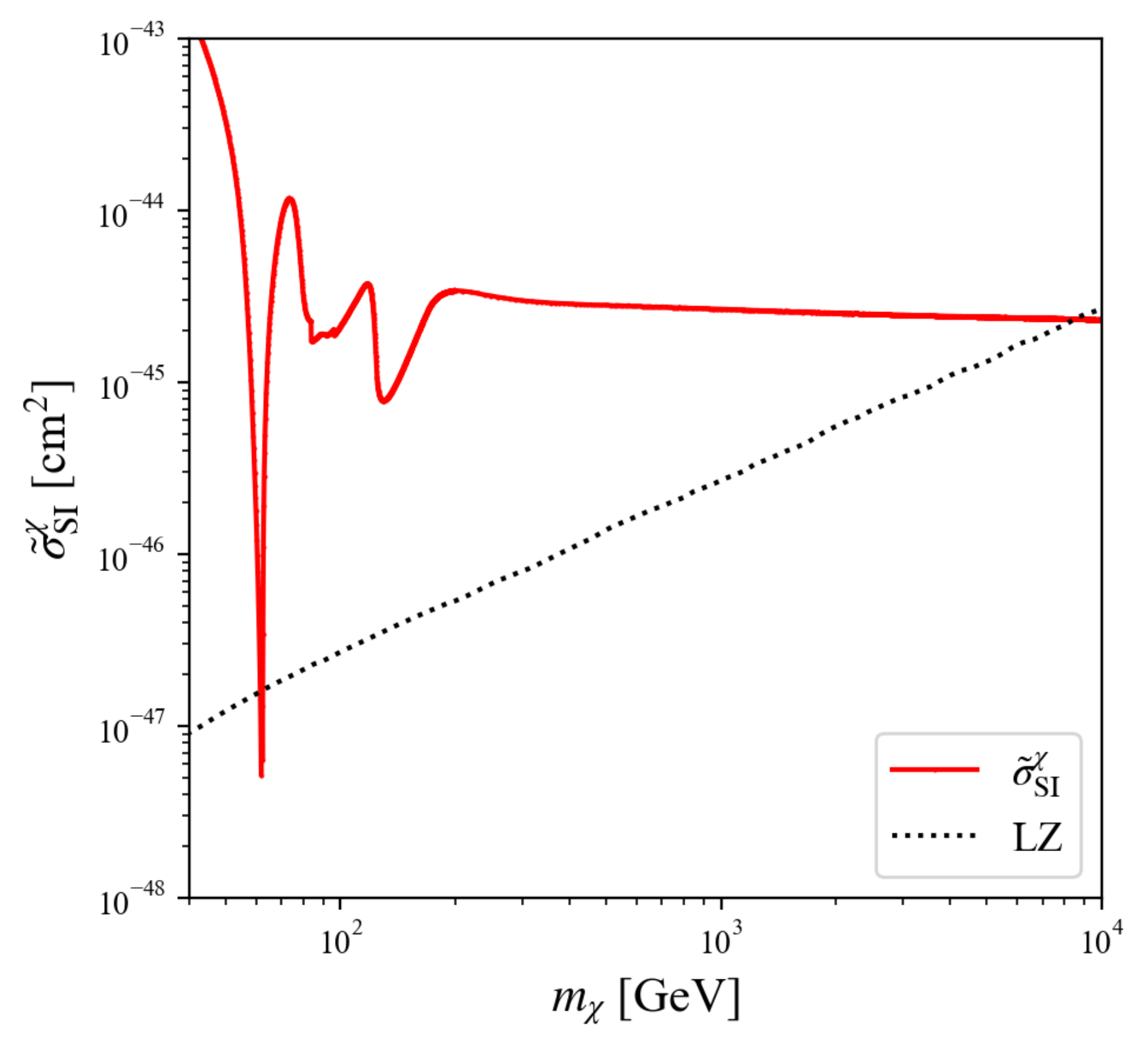}
\caption{DM relic density $\Omega_\chi h^2$ (left panel) and scattering cross section with nucleons $\tilde{\sigma}_\mathrm{SI}^\chi$ (right panel) plotted against DM mass $m_\chi$. BP1 in Table~\ref{tab:BP} is used input parameters other than $m_\chi$. The dotted line in the left panel represents the central value of the DM relic density, and the dotted line in the right panel represents the LZ data.}
\label{fig:chinum}
\end{center}
\end{figure}

In Fig.~\ref{fig:chinum}, we display DM relic density $\Omega_\chi h^2$ and scattering cross section with nucleons $\tilde{\sigma}_\mathrm{SI}^\chi$ as a function of DM mass $m_\chi$.  
For inputs other than $m_\chi$, we adopt input parameters of BP1 in Table~\ref{tab:BP}, which realizes the strong first-order phase transition. Here, when $\Omega_\chi h^2$ is less than the observed value, we scale $\sigma_\mathrm{SI}^\chi$ as
\begin{align}
\tilde{\sigma}_{\mathrm{SI}}^\chi=\left(\frac{\Omega_\chi}{\Omega_{\mathrm{DM}}}\right) \sigma_{\mathrm{SI}}^\chi,
\end{align}
where the observed value is~\cite{Planck:2018vyg}
\begin{align}
\Omega_{\text{DM}}h^2 = 0.1200\pm 0.0012.
\label{Oh2obs}
\end{align}
It can be seen that only at $m_\chi \approx 62.5~\mathrm{GeV}$ the DM relic density and the LZ results are satisfied
\footnote{This is un update of the discussion in ref.~\cite{Cho:2021itv} based on the XENON1T experiment~\cite{XENON:2020kmp}. 
The DM mass $m_\chi\simeq 62.5~\mathrm{GeV}$ is still allowed, but $m_\chi \simeq 2~\mathrm{TeV}$ in the previous analysis is excluded by this update of the results of the direct detection experiments.}.

However, since the DM mass $m_\chi$ is almost half of the mass of degenerate scalars, $m_{h_1} \approx m_{h_2} \approx 125~\mathrm{GeV}$, 
the relic density of the scalar DM is much smaller than the observed value, 
due to the resonant enhancement of the DM annihilation process mediated by $h_1$ and $h_2$.
%\footnote{As mentioned at the beginning of Sec.~\ref{sec:dege}, the scattering of scalar DM $\chi$ and nucleons is suppressed by the degenerate scalar scenario. Quantitative results are shown in ref.~\cite{Cho:2021itv}.}
The scaled cross section $\tilde{\sigma}_\mathrm{SI}^\chi$ is directly affected and suppressed by this effect.
Therefore, in the following numerical study, we fix the scalar DM mass at $m_\chi = 62.5~\mathrm{GeV}$ with $m_{h_1} \approx m_{h_2}$ to satisfy the condition realizing the first-order EWPT and constraints from the direct detection experiments.  
Then, we investigate the possibility of the fermion DM compensating for the lack of thermal relic abundance in the scalar DM.

\begin{table}[t]
\center
\begin{tabular}{|c|c|c|c|c|c|c|c|}
\hline
Inputs & $v$ [GeV] & $m_{h_1}$ [GeV] & $m_{h_2}$ [GeV] & $\alpha$ [rad] & $a_1$ [GeV$^3$] & $v_S$ [GeV] & $m_\chi$ [GeV]  \\ \hline
 BP1 & 246.22 &125 & 124 & $\pi/4$ & $-6576.17$ & 0.60 & 62.5  \\ \hline
 BP2 & 246.22 &125 & 124 & $\pi/8$ & $-4904.17$ & 0.45 & 62.5  \\ \hline
 BP3 & 246.22 &125 & 124 & $\pi/16$ & $-2719.786$ & 0.25 & 62.5  \\ \hline
 BP4 & 246.22 &125 & 124 & $0.045$ & $-652.36971$ & 0.060 & 62.5  \\ \hline

Outputs & $m^2$ [GeV$^2$] & $b_1$ [GeV$^2$] & $b_2$ [GeV$^2$] & $\lambda$ & $a_1$ [GeV$^3$] & $d_2$ & $\delta_2$  \\ \hline
 BP1 & $-(124.5)^2$ &$(107.7)^2$ & $-(178.0)^2$ & 0.511 & $-6576.17$ & 1.77 & 1.69  \\ \hline
 BP2 & $-(124.9)^2$ &$(107.3)^2$ & $-(169.9)^2$ & 0.514 & $-4904.17$ & 1.44 & 1.59  \\ \hline
 BP3 & $-(125.0)^2$ &$(107.1)^2$ & $-(166.2)^2$ & 0.515 & $-2719.786$ & 1.41 & 1.55  \\ \hline
 BP4 & $-(125.0)^2$ &$(107.1)^2$ & $-(163.2)^2$ & 0.515 & $-652.36971$ & 1.31 & 1.51  \\ \hline
 
\end{tabular}
\caption{Input and output parameters for strong first-order EWPT. The two independent parameters $m_f$ and $\lambda_f$ associated with the fermion DM are treated as variables.}
\label{tab:BP}
\end{table}

We adopt a set of input parameters as shown in Table~\ref{tab:BP}.
As mentioned above, a sizable $\delta_2$ is required for a strong first-order EWPT to occur, the typical magnitude of which is $\delta_2 \sim 1$, realized by a relatively large $\alpha$ and small $v_S$. Even if $\alpha$ is reduced from the maximum value $\pi/4$, $v_S$ keeps the magnitude of $\mathcal{O}(0.1)$ for a while~(BP1-BP3). On the other hand, when $\alpha\lesssim 0.1$, the appropriate magnitude of $v_S$ is $\mathcal{O}(0.01)$~(BP4).
Output parameters obtained through \eqref{m2}-\eqref{b1} are also shown in the table. The following analysis treats the fermion DM mass $m_f$ and the Yukawa coupling $\lambda_f$ as free parameters with these inputs. 

\begin{figure}[htpb]
\begin{center}
\includegraphics[width=7cm]{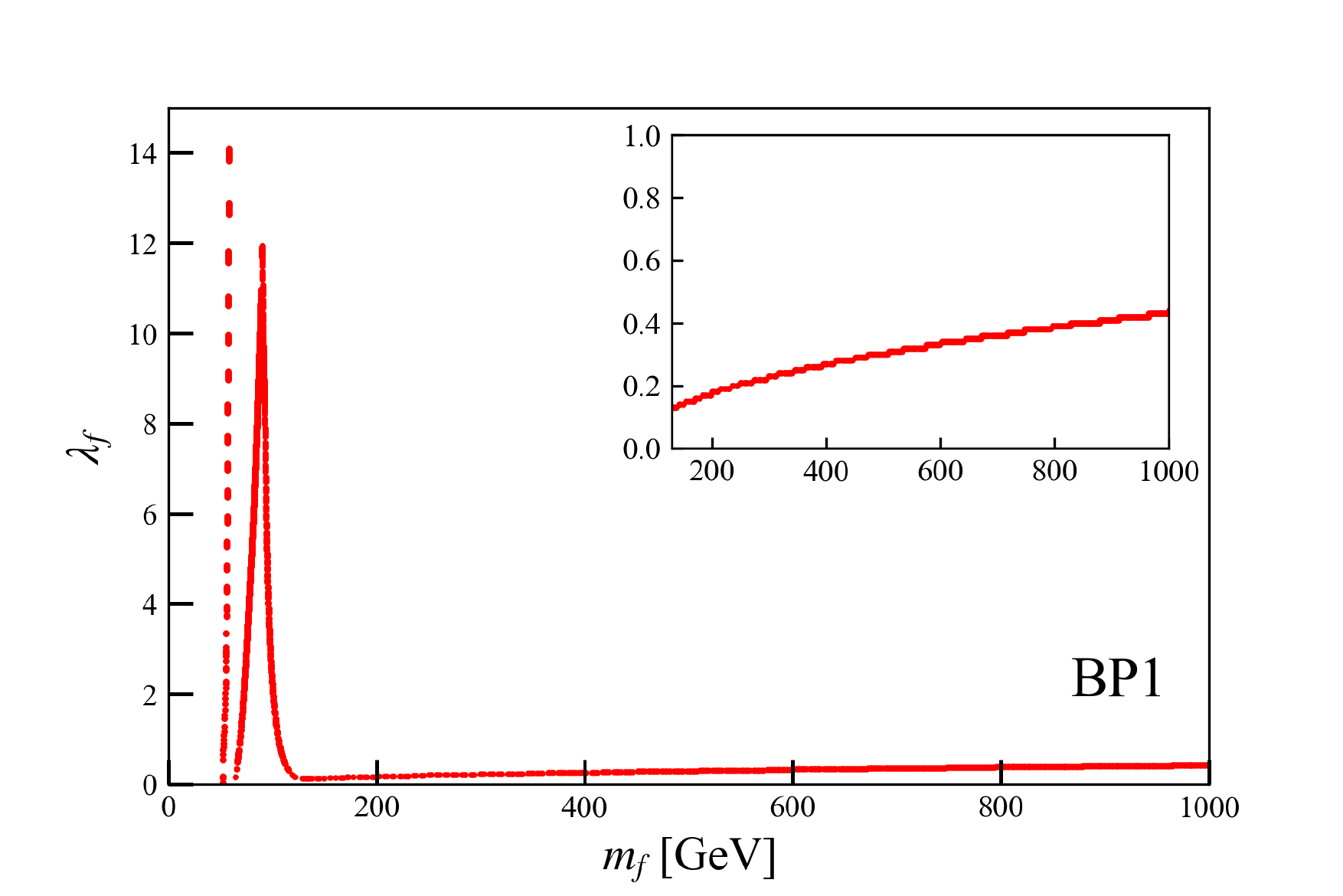}
\includegraphics[width=7cm]{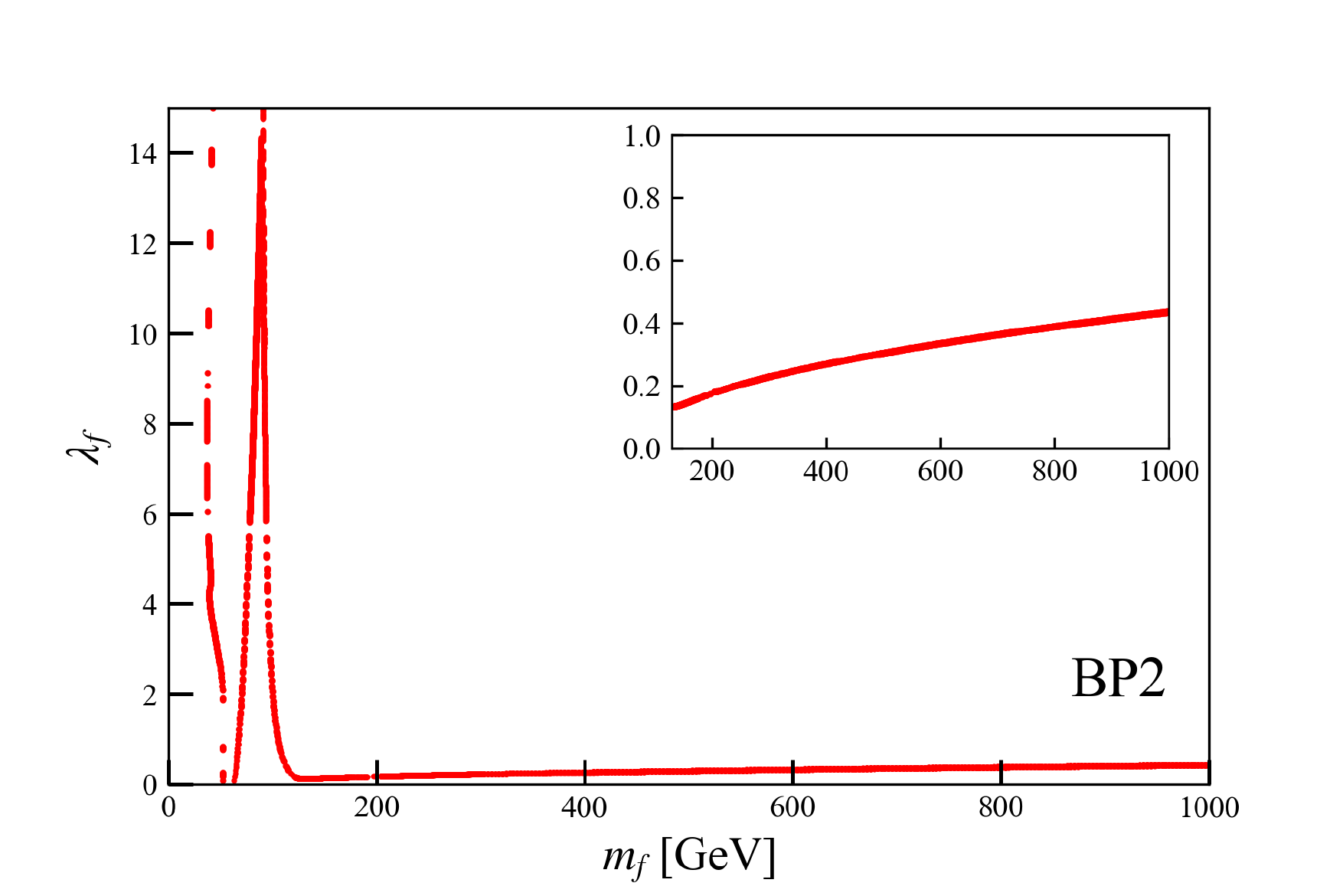}
\includegraphics[width=7cm]{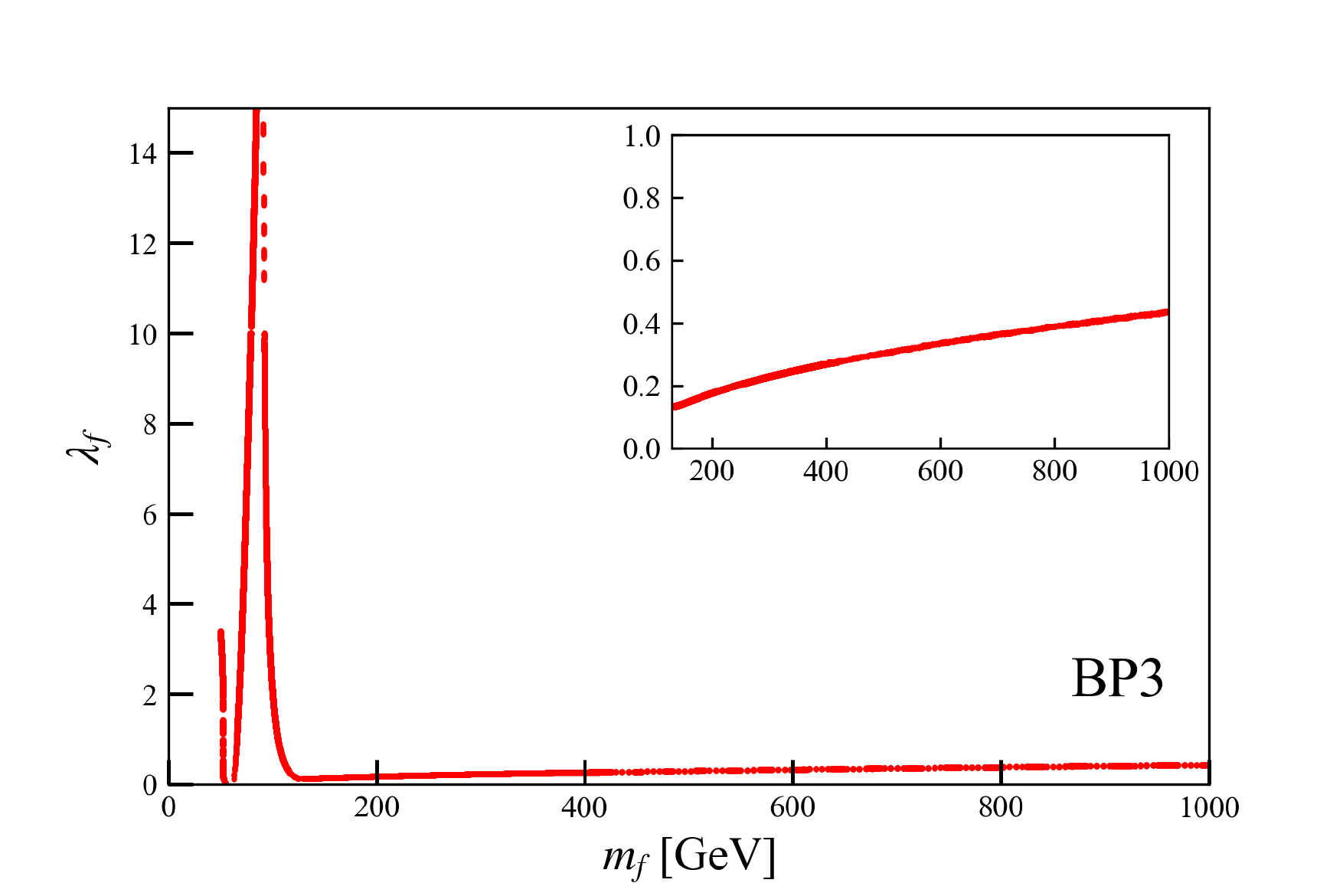}
\includegraphics[width=7cm]{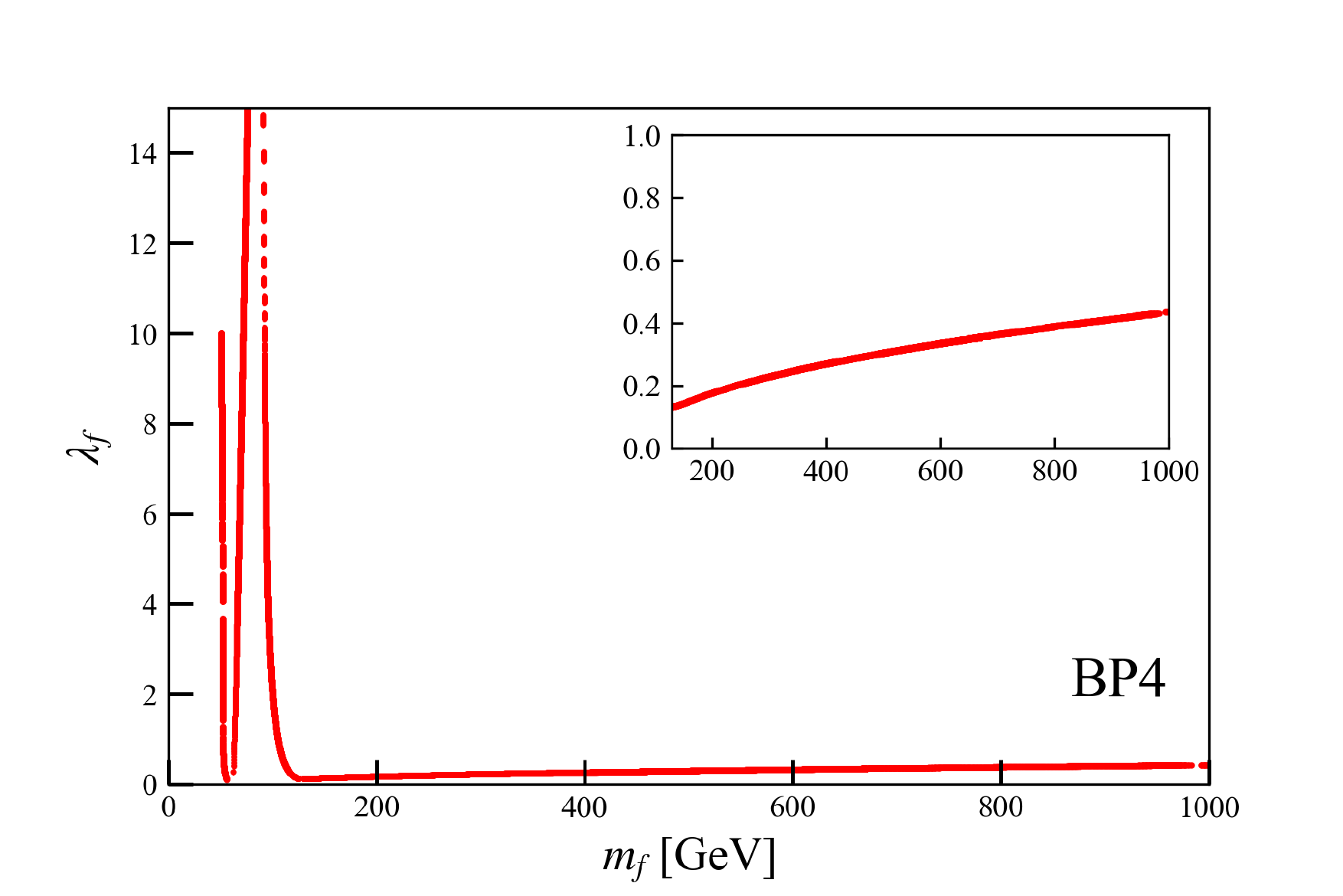}
\caption{Scatter plot of $\qty(m_f,~\lambda_f)$ satisfying the observed relic density~\eqref{Oh2obs}  at the 2-$\sigma$ level for BP1-BP4.}
\label{fig;reliccontour}
\end{center}
\end{figure}

Since there are two DM candidates in our model, 
the sum of the relic density of the scalar DM $\qty(\Omega_\chi h^2)$ and of the fermion DM $\qty(\Omega_f h^2)$, i.e.,  
\begin{align}
\Omega_\text{total} h^2= \Omega_\chi h^2+\Omega_f h^2. 
\end{align}
should be compared with the observed value~\eqref{Oh2obs}.
We show the allowed combinations of $\qty(m_f,~\lambda_f)$ from the observed relic density \eqref{Oh2obs} at the 2-$\sigma$ level in Fig.~\ref{fig;reliccontour} and the small boxes in the figures enlarge the trend of 
allowed $\qty(m_f,~\lambda_f)$ for $m_f > 130~\mathrm{GeV}$.
It can be seen that similar results are found with any benchmark point.
Note that the relic density of the scalar DM $\Omega_\chi h^2$ is negligibly tiny due to resonant enhancement of annihilation processes mediated by $h_1$ and $h_2$, owing to the scalar DM with $m_\chi=62.5~\mathrm{GeV} \approx m_{h_1}/2~\qty(\approx m_{h_2}/2)$. 
Thus, the fermion DM accounts for most of the total DM.  
It can be seen from Fig.~\ref{fig;reliccontour} that the Yukawa coupling $\lambda_f$ is allowed to be $\mathcal{O}(0.1) - \mathcal{O}(10)$ at relatively small DM mass, $m_f \lesssim 125~\mathrm{GeV}$. This is because the relic density varies drastically in the range of $m_f$. 
At $m_f \approx 62.5~\mathrm{GeV}$, the relic density $\Omega_f h^2$ is decreased significantly as compared to smaller $m_f$ due to the resonant annihilation of the fermion DM pair mediated by the Higgs bosons. 

In addition, at $m_f \approx 125~\mathrm{GeV}$, 
the annihilation channel into the Higgs pair opens kinematically. 
Since this process is not suppressed with the degenerate scalar scenario, it occurs frequently and decreases the DM abundance. 
Thus, for $m_f \approx 62.5~\mathrm{GeV}$ and $m_f \gtrsim 125~\mathrm{GeV}$, small $\lambda_f$ is preferred to avoid excessive DM annihilation.
On the other hand, in the region, 62.5 GeV $\lesssim m_f \lesssim$ 125 GeV, the annihilation amplitudes of the fermion DM into the SM particles are suppressed due to degenerate scalar mediators. However, since the mass difference of two scalars is fixed at $m_{h_1}-m_{h_2}=1$ GeV, the cancellation of the amplitudes is incomplete. Therefore, the Yukawa coupling $\lambda_f$ varies with $m_f$, and it peaks at $m_f \approx 90~\mathrm{GeV}$, right in the middle of the region ($62.5~\mathrm{GeV}\lesssim m_f \lesssim 125~\mathrm{GeV}$)\footnote{The position of the peak depends on the input parameters. For example, the peak appears at $m_f\approx 100$ GeV when the mass difference $m_{h_1}-m_{h_2}=1.1$ GeV.}. 

\begin{figure}[htpb]
\begin{center}
\includegraphics[width=7cm]{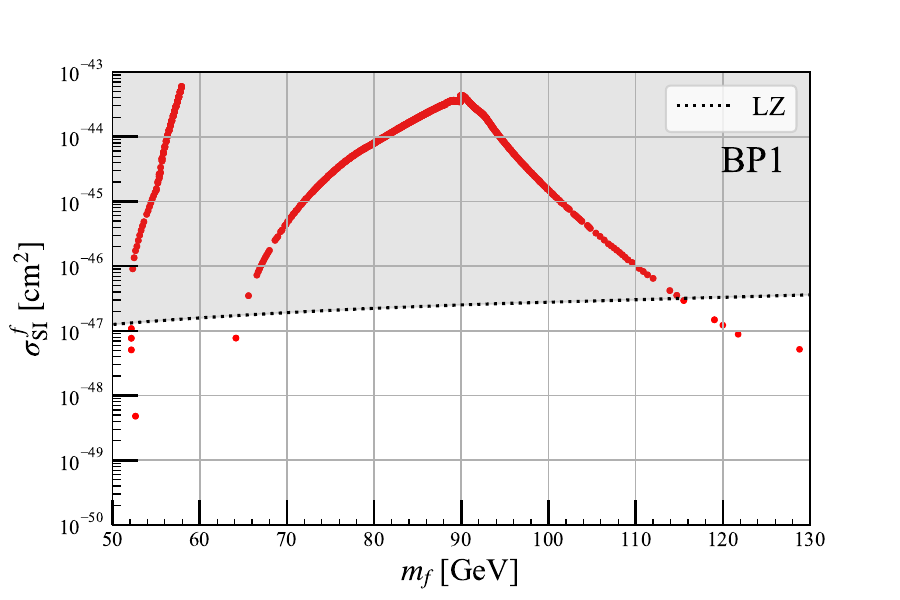}
\includegraphics[width=7cm]{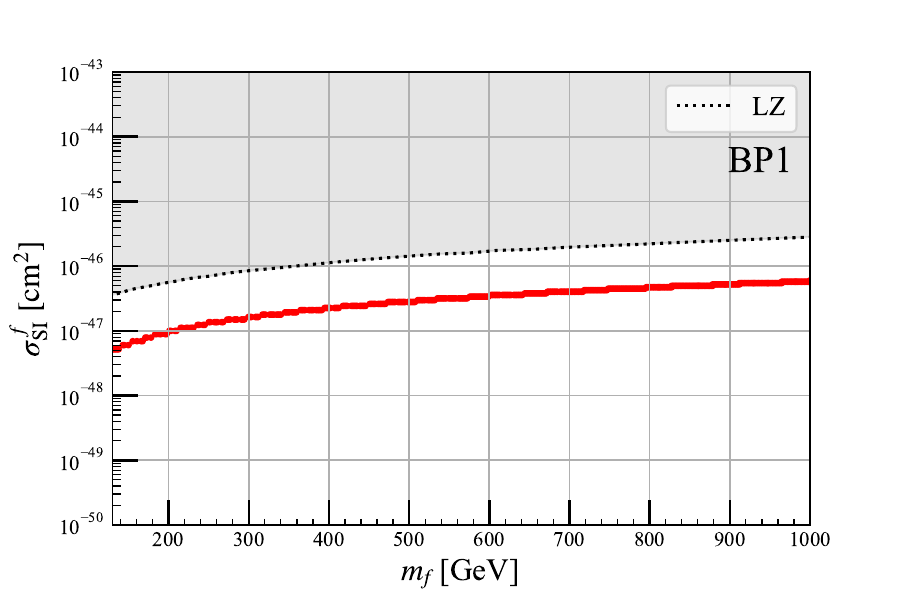}
\includegraphics[width=7cm]{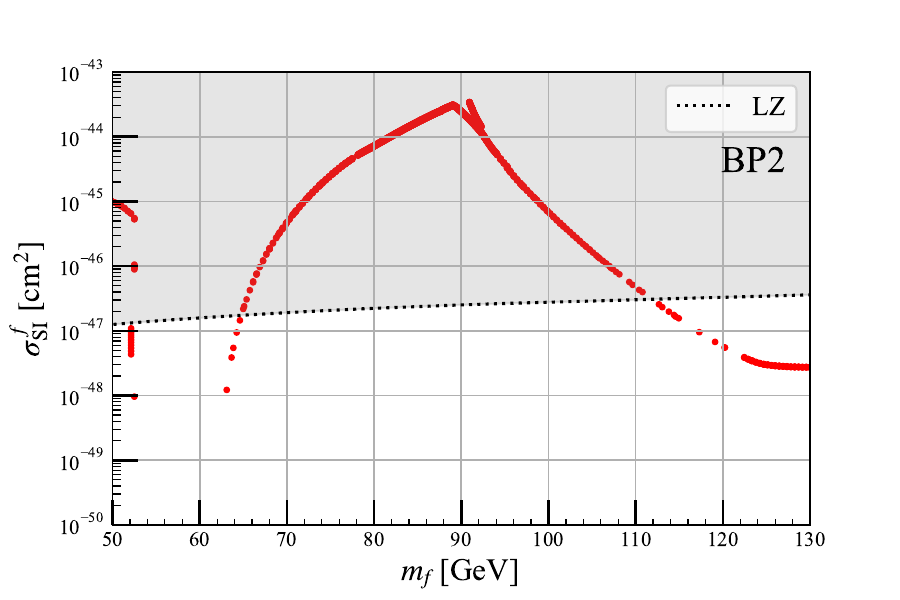}
\includegraphics[width=7cm]{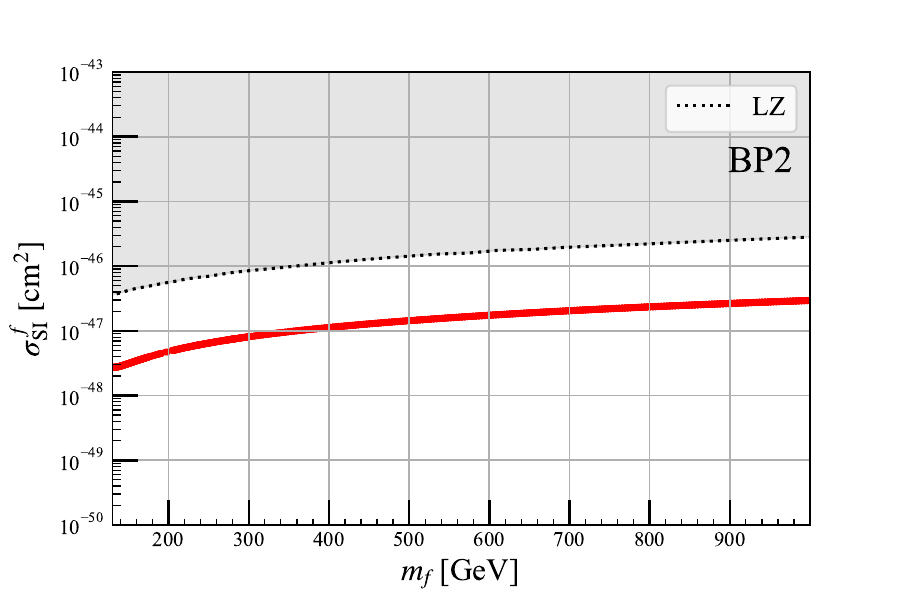}
\includegraphics[width=7cm]{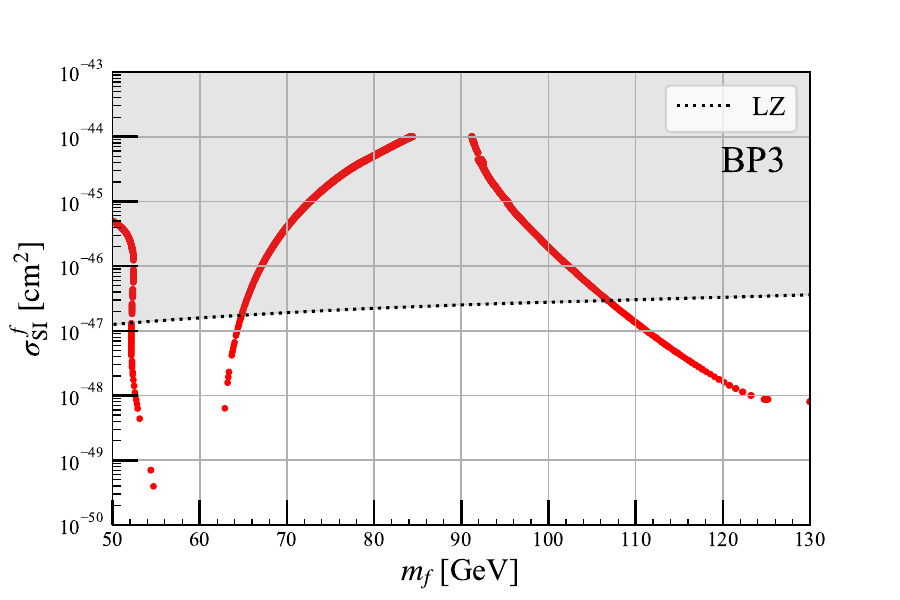}
\includegraphics[width=7cm]{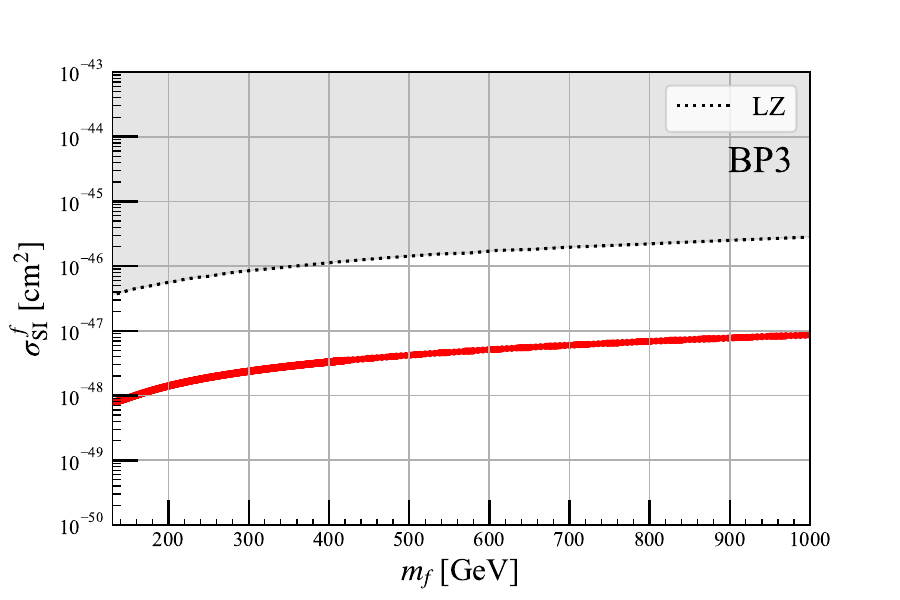}
\includegraphics[width=7cm]{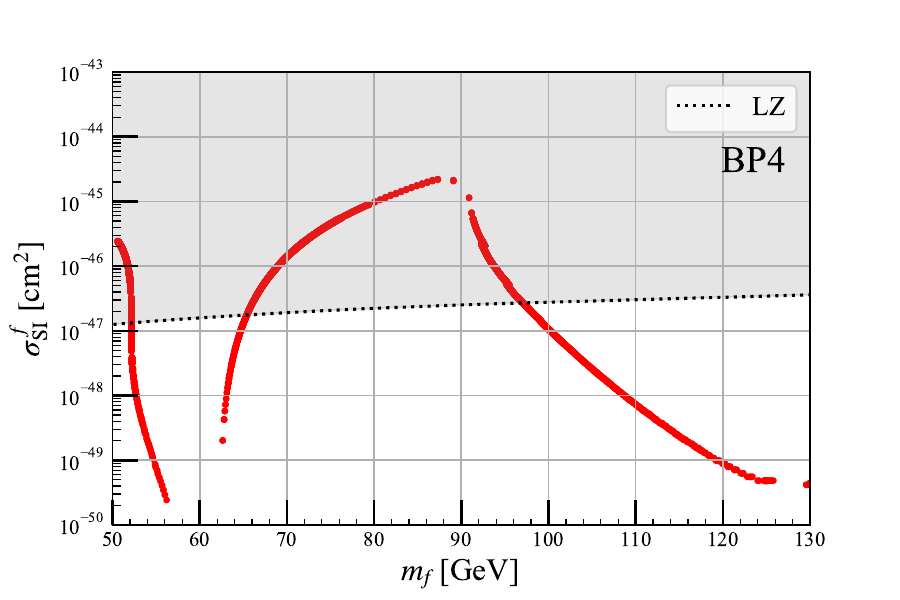}
\includegraphics[width=7cm]{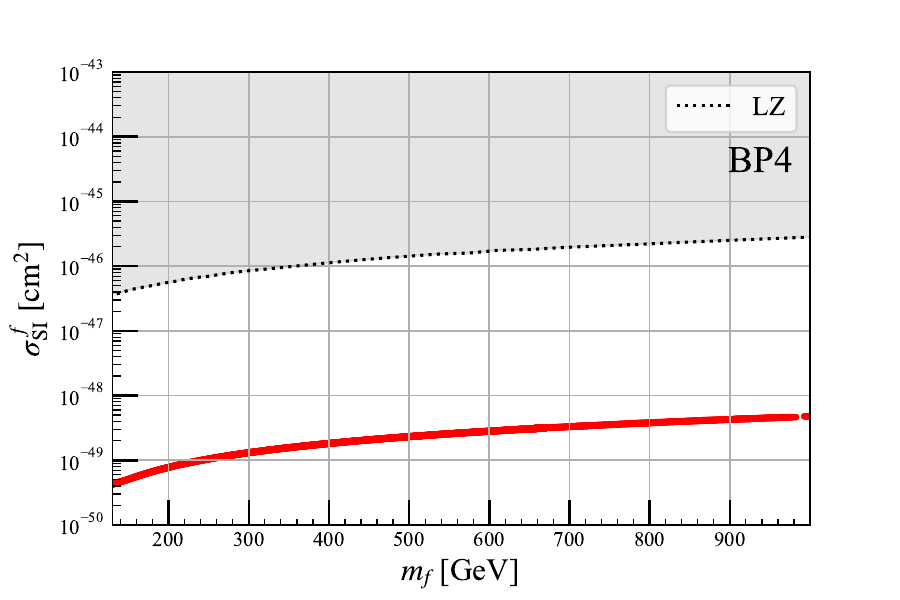}
\caption{Spin-independent scattering cross section between the fermion DM and nucleons $\sigma_\mathrm{SI}^f$ as a function of $m_f$ for BP1 - BP4. All points shown reflect the points of Fig.~\ref{fig;reliccontour} and satisfy the observed relic density~\eqref{Oh2obs} at the 2-$\sigma$ level. The dotted line represents the results of the LZ experiment~\cite{LZ:2022ufs}. The left graph corresponds to the region $m_f<130$ GeV and the right graph to the region $130<m_f<1000$ GeV.}
\label{fig;sigma}
\end{center}
\end{figure}

The spin-independent cross sections between the fermion DM $f$ and nucleons $\sigma_\mathrm{SI}^f$ obtained by each set of allowed ($m_f,~\lambda_f$) in Fig.~\ref{fig;reliccontour} is projected onto $m_f$-$\sigma_\mathrm{SI}^f$ plane in Fig.~\ref{fig;sigma}. 
The black-dotted line represents the upper bound on the cross section at 95\% CL given by the LZ experiment~\cite{LZ:2022ufs}. 
Note that we focus only on the fermion DM here since contributions to the scattering process by the scalar DM are negligibly small owing to the suppression mechanism on the process due to the degenerate scalars as described in Sec.~\ref{sec:dege}. The four left panels are for $m_f < 130~\mathrm{GeV}$ while the four right ones are for $130~\mathrm{GeV} < m_f< 1000~\mathrm{GeV}$ which correspond to the small boxes in Fig.~\ref{fig;reliccontour}. 
We find that the exclusion of the parameters $(m_f,~\lambda_f)$ allowed from the DM relic density shown in Fig.~\ref{fig;reliccontour} can be found only for $m_f < 130~\mathrm{GeV}$ (left panels), while all points in the right panels satisfy the bounds from the LZ experiment. However, compared to BP1-3, BP4 gives a cross section about two orders of magnitude smaller, and we describe how this is qualitatively explained.
The cross section $\sigma_{\mathrm{SI}}^f$ is given by 
\begin{align}
\sigma_{\mathrm{SI}}^f 
&\propto 
\lambda_f^2 \sin ^2 \alpha \cos ^2 \alpha\left(\frac{1}{m_{h_1}^2}-\frac{1}{m_{h_2}^2}\right)^2 
\nonumber \\
&=
\frac{\lambda_f^2 v^2}{4 m_{h_1}^4 m_{h_2}^4} \delta_2^2 v_S^2.
\label{fermionsigma}
\end{align}
In contrast to the cross section of the scalar DM $\sigma_\mathrm{SI}^\chi$ \eqref{scalarsigma}, the singlet VEV $v_S$ is present in the numerator in \eqref{fermionsigma}. Since $v_S$ is chosen to be $\mathcal{O}(0.1)$ GeV for BP1-3 and $\mathcal{O}(0.01)$ GeV for BP4 in our numerical study as required from strong first-order EWPT, the cross section $\sigma_{\mathrm{SI}}^f$ for BP4 is directly affected by the size of $v_S$ and becomes smaller than those for the other benchmark points. 
Let us remind the reader that the Yukawa coupling $\lambda_f$ is chosen to satisfy constraints from the DM relic density. 
Fig.~\ref{fig;reliccontour} tells us that $\lambda_f \sim \mathcal{O}(10)$ for $m_f \sim 90~\mathrm{GeV}$, while $\lambda_f \sim \mathcal{O}(0.1)$ for smaller or larger $m_f$. 
As shown in \eqref{fermionsigma}, since $\sigma_{\mathrm{SI}}^f$
is proportional to $\lambda_f^2$, this explains the difference of magnitude of $\sigma_{\mathrm{SI}}^f$ between at $m_f \sim 90~\mathrm{GeV}$ and at the other value of $m_f$ in Fig.~\ref{fig;sigma}. 
What can be said through all benchmark points is that $\sigma_{\mathrm{SI}}^f$ is suppressed except for $m_f < 130~\mathrm{GeV}$ where the Yukawa coupling $\lambda_f$ is sizable (see, Fig.~\ref{fig;reliccontour}). The insufficient relic density in the CxSM with the degenerate scalar scenario, therefore, could be compensated by the fermion DM.

\begin{figure}[htpb]
\begin{center}
\includegraphics[width=9cm]{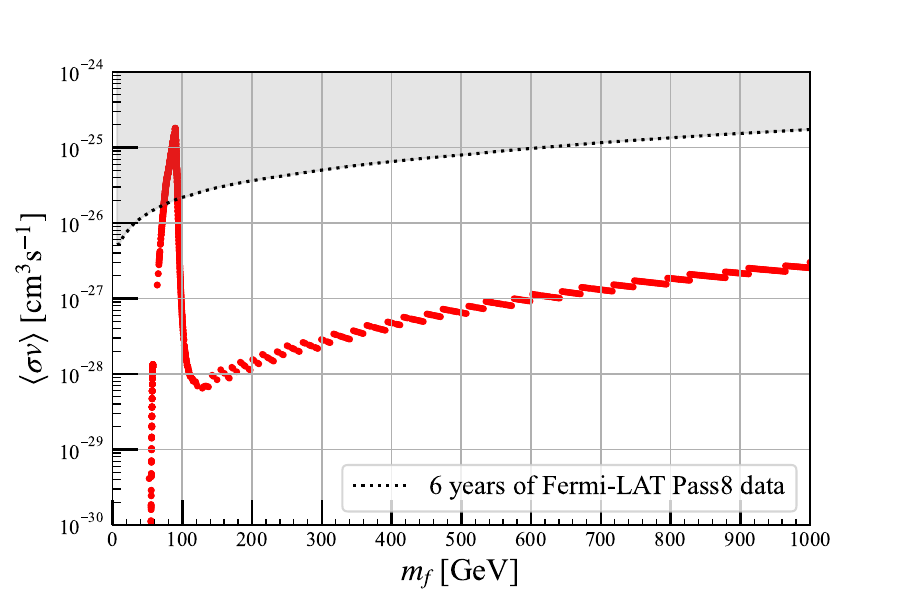}
\caption{The DM annihilation cross section as a function of $m_f$ for BP1. The dotted line represents the results of the fermi-LAT gamma-ray observations~\cite{Fermi-LAT:2015att}.}
\label{indirect}
\end{center}
\end{figure}

Finally, we mention constraints on the DM annihilation cross section from indirect detection experiments. In Fig.~\ref{indirect}, the DM annihilation cross section $\langle \sigma v \rangle$ for fermion DM mass $m_f$ is shown, where BP1 is adopted as an example. 
The Yukawa coupling $\lambda_f$ is chosen to reproduce the observed DM relic density; see Fig.~\ref{fig;reliccontour}. 
The dotted line represents data from fermi-LAT gamma-ray observations of the dwarf spheroidal satellite galaxies of the Milky Way~\cite{Fermi-LAT:2015att}. 

The annihilation rates of DM pairs into quarks, leptons, and gauge bosons are suppressed by the cancellation mechanism of the degenerate scalar scenario so that these processes are not severely constrained from indirect detection experiments in many mass regions.
In fact, as can be seen from Fig.~\ref{indirect}, the annihilation cross section in $m_f\lesssim 62.5~\rm{GeV}$ is very suppressed since the DM pair annihilates into the SM particles except for $W, Z$ bosons, top quark and Higgs bosons. 
However, we emphasize that the annihilation processes from fermion DM to scalar $ff\to\chi\chi$ ($m_f\gtrsim 62.5~\rm{GeV}$), $~ff\to h_ih_j~(i,j=1,2)$ ($m_f\gtrsim 125~\rm{GeV}$) are not suppressed even with the degenerate scalar scenario. 
For $62.5~\mathrm{GeV} \lesssim m_f \lesssim 125~\mathrm{GeV}$, since the main annihilation mode is $ff\to\chi\chi$ while the channel $ff\to h_ih_j$ is kinematically forbidden, the Yukawa coupling $\lambda_f$ should be as large as $\mathcal{O}(10)$ to reproduce the observed relic density~\eqref{Oh2obs} as shown in Fig.~\ref{fig;reliccontour}.
On the other hand, for $m_f \gtrsim 125~\mathrm{GeV}$, the cross section gradually increases. 
Since, in this region, the Yukawa coupling $\lambda_f$ is required to be $\mathcal{O}(0.1)$ to achieve the observed relic density, the annihilation cross section is sufficiently smaller than the limit from the Fermi LAT experiment. We conclude that no further constraints on the model parameter space are obtained from the indirect detection experiments. 
Since the processes $ff\to\chi\chi$ and $ff\to h_ih_j$ are $p$-wave annihilation processes, the annihilation cross section is suppressed as seen from almost all the regions of $m_f$ in Fig.~\ref{indirect}.
Even though the region excluded by the Fermi LAT experiment in Fig.~\ref{indirect} reflects the contribution of the annihilation channel $ff\to\chi\chi$, $\chi$ does not produce the gamma-rays so that the constraints on the annihilation cross section would be relaxed to some extent.

\section{Summary}\label{sec:sum}

We have investigated the possibility of solving the difficulty of the CxSM with the degenerate scalar scenario by adding an extra singlet fermion. 
It has been pointed out that although the CxSM with degenerate scalar scenario can realize strong first-order EWPT without suffering from the constraints from the direct detection experiments, the relic density of the scalar DM in the model is too tiny to explain the observed relic density of DM. 
This fact, the insufficient relic density by the scalar DM, requires a new DM candidate. 
In the CxSM, the imaginary part of the singlet scalar behaves as the scalar DM due to the CP symmetry of the scalar potential. 
The scattering amplitudes of the scalar DM and quarks mediated by Higgs bosons are known to be suppressed when masses of two Higgs bosons are degenerate. However, the condition of the model parameters for the degenerate scalar scenario conflicts with the condition for realizing strong first-order EWPT. 

We have shown that using the benchmark point where strong first-order EWPT is achieved, the LZ experiment gives strict bounds on the cross section, though many DM mass regions predict small relic abundance. 
As a result, it is found that the model satisfies both constraints from the direct detection experiments and the condition of strong first-order EWPT only when $m_\chi=62.5~\mathrm{GeV}$ (see, Fig.~\ref{fig:chinum}).

As a price of simultaneous fulfillment of two conflicting conditions, the relic density of the scalar DM is highly suppressed, reflecting the resonance enhancement of the DM annihilation process via the Higgs exchange. 
Therefore, the tiny relic density of the scalar DM requires introducing a new DM candidate.

In this study, we have introduced a vector-like fermion, which only couples to the singlet scalar $S$, as a new DM candidate. 
We found that the fermion DM-quark scattering mediated by two degenerate scalars $(h_1,~h_2)$ was suppressed due to the orthogonality of the mixing matrix, which defines the mass eigenstate, as in the case of the scalar DM. 
Since the mass $m_f$ and the Yukawa coupling $\lambda_f$ of the fermion DM are independent of the scalar potential in the CxSM, i.e., of the condition of strong first-order EWPT, we focused on contributions of the fermion DM to the scattering with quark and the relic density. 
We have examined the favored combination of $(m_f,\lambda_f)$, which is consistent with the observed relic density of the DM at the 2-$\sigma$ level. 
For relatively light fermion DM ($m_f \lesssim 130~\mathrm{GeV}$), the Yukawa coupling $\lambda_f$ could be $\mathcal{O}(1)$-$\mathcal{O}(10)$, while 
$\lambda_f \lesssim 0.5$ is required for heavier fermion DM ($130~\mathrm{GeV} \lesssim m_f \lesssim 1000~\mathrm{GeV}$). 
The large Yukawa coupling $\lambda_f$ increases the cross section of the fermion DM-quark scattering even under the degenerate scalar scenario, so an upper limit is placed on $\lambda_f$. 
So, some range in $m_f \lesssim 130~\mathrm{GeV}$ is disfavored from the direct detection experiments, while the larger $m_f$ 
($\gtrsim 130~\mathrm{GeV}$) is less constrained. 
Nevertheless, compared to the scalar DM studied in ref.~\cite{Cho:2021itv}, the allowed parameter space has been enlarged by introducing the fermion DM. Therefore, the current DM experiments and/or observations can be well explained by the scalar DM and the fermion DM while respecting strong first-order EWPT realization.

\begin{acknowledgments}
The work of GCC is supported in part by JSPS KAKENHI Grant No. 22K03616. The work of CI is supported by JST, the establishment of university fellowships towards the creation of science and technology innovation, Grant No. JPMJFS2113.

\end{acknowledgments}

\bibliography{biblist}

\end{document}